# ShielDroid: A Hybrid Approach Integrating Machine and Deep Learning for Android Malware Detection

Md Faisal Ahmed, Zarin Tasnim Biash , Abu Raihan Shakil ,Ahmed Ann Noor Ryen,Arman Hossain,Faisal Bin Ashraf , Muhammad Iqbal Hossain
Department of Computer Science and Engineering Brac University
Dhaka, Bangladesh

***Abstract*—The rapid advancement of modern technology has led to a significant increase in the use of smart devices, such as smartphones and tablets, resulting in the widespread adoption of mobile applications. Although applications are required to undergo malware screening before being published on official app stores, many malicious applications successfully evade detection by concealing sophisticated malware variants. These malicious behaviors are often activated only during runtime, making them difficult to identify through conventional static analysis. As a result, malware may remain undetected until after installation, potentially causing irreversible damage to users and their devices. This study presents a real-time Android malware detection framework that analyzes application behavior to accurately identify and classify complex malware. The proposed approach employs a hybrid dynamic analysis technique to distinguish malicious applications from benign ones. After preprocessing and filtering the collected dataset, the applications are classified using multiple machine learning algorithms. A comprehensive performance evaluation is conducted to compare the effectiveness of different classification techniques in terms of detection accuracy and execution time. Experimental results demonstrate that a hybrid model combining Random Forest and a Multilayer Perceptron achieves the best overall performance, attaining an accuracy of 97.5% with an execution time of 22.945 seconds. The proposed framework can enhance mobile device security by enabling timely detection of malicious applications and reducing the risk of cyberattacks.**

## I. Introduction

In recent years, smartphones have become widely available and accessible to individuals all around the world giving rise to security concerns. According to the research, the methodology for uploading an app to the Android app market is less restrictive than iOS's App Store [1]. Hackers have been using Google Play for years to deploy an extraordinarily complex backdoor capable of collecting a wide variety of sensitive data. A large number of Android applications are being developed and submitted to the Play Store. As a result, an efficient malware detection system for detecting malware in apps published to the Play Store is important. According to the article[2], Android malware is similar to the many varieties of malware seen on desktop and laptop computers targeting Android phones and tablets. Malicious software or code meant to harm a user's device, such as trojans, adware, ransomware, spyware, viruses, or phishing applications, is referred to as mobile malware. Researchers and developers use static analysis, dynamic analysis, and artificial intelligence to assess various security solutions to prevent malware attacks. More than 140 million new malware samples have been found according to recent statistics in 2015 [3]. In a 2019 survey done by researchers at Check Point, they found that the rate of cyber-attacks in smartphones and other devices has risen by 50% than that of the last year [4]. Many applications successfully pass the first screening test during the accepting procedure despite having malicious contents due to the malware being difficult to distinguish. In such circumstances, users discover that the application contains malware after it has been installed and executed when it might be too late to undo the malware's damage. Therefore, a strong real-time Android malware detection to swiftly analyze malicious contents which are difficult to detect, before the malware can do any damage is the focus of this research. The CICMaldroid 2020 dataset has been used for this work which is relatively new and consists of the latest malware samples that could be successfully detected. Furthermore, after studying several papers, it has been found that the other works use basic classification models, unlike this work where a hybrid model has been used for better performance. Moreover, the work is robust on high-dimensional data, which allows it to work effectively when used with a new and similar independent dataset. In addition, a prototype application has been demonstrated in this research that has several scopes of improvement in the future, e.g. it can be made more dynamic and functioning with real-time data.

## II. Related work

Researchers have been particularly concerned about Android malware, and as the number of Android users has expanded, malware analysis has become more intensive. According to [5], in 2018, the number of Android smartphone users in the US rose by 120.5 million, with a prediction of 130 million by the end of 2021. Moreover, the article [6] states, in the mid-2017, malware known as 'Trojan Virus' was hidden in the sight of a game in Google Play Store and more than 50000 people downloaded it thinking it was a game. This section covers the ways of detecting Android malware as well as the challenges that come with it.

The research paper [7] describes some of the most widely used network datasets in Machine Learning and Deep Learning and discusses the three types of intrusion detection systems (IDSs) available: signature-based or misuse-based, anomaly-based, and hybrid. The analysis [8] discusses the application and frequency-based attacks on mobile phones. The authors used a software-based radio circuit to successfully attack an Android phone in an experiment. Another recent research [9] suggests detecting malicious applications based on the application's permission request and API call frequency distribution. Then, using 10-fold cross-validation, machine learning methods were used, yielding an F-measure of 94.3%. According to the results of the experiment, the random tree and k-NN algorithms are the fastest for training and testing, whereas J48 takes the most time. This systematic study [10] developed a hybrid model incorporating Deep Encoder (DAE) and convolutional neural network to improve the accuracy and efficiency of Android virus detection (CNN). With varied topologies, CNN was able to obtain a detection accuracy of about 99.8%, which is 5% higher than SVM. However, it took a long time to attain this level of precision (332.9 minutes). This recent paper [1] proposes a sequence-based malware detection system called 'SeqMobile,' which accepts three effective performance optimization strategies to detect malware with a 97.85% accuracy and meet the demand for Android device run-time performance.

From the above discussion, a brief summarization of different approaches for developing an Android malware detection system can be achieved.

## III. Proposed Methodology

The proposed methodology begins with data preparation and then selects related features from a large number of options. Random Forest classification is used for generating a predicted column by building several decision trees. Finally, a feed-forward neural network is trained using the predicted column as an additional input.

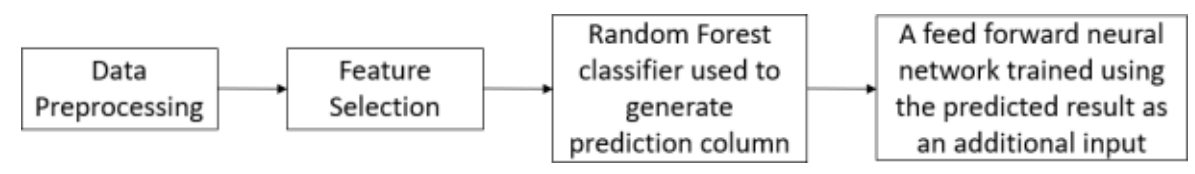

Fig. 1. Workflow diagram of the proposed methodology

### A. Data Preprocessing

Data preprocessing is a data mining approach for converting raw data into a valuable and efficient format. Normalization is done, and noisy data is taken into account to eliminate it if it is discovered. Data preprocessing is performed through standardization by calling the StandardScaler method using sklearn. The data from the dataset has been screened for noise. But no noise was found, due to which no noisy data removal technique has been performed.

### B. Feature Selection

For extracting the characteristics for this work, Analysis of Variance has been used. The Selectkbest function in the scikit-learn library uses the ANOVA (Analysis of variance) test. ANOVA test figures out the difference between testing groups. It is a t-test generalization for more than two groups. The independent t-test is then used to compare two groups' means or averages for a condition [11].

### C. Model Description

Random Forest classifier is used to generate a predicted column by constructing numerous decision trees. Finally, to increase overall accuracy, the anticipated column is utilized as an extra input to train a feed-forward neural network. When designing the model, it is made sure to limit the training time to a minimum.

*1) Random Forest Classifier:* To begin, the dataset was divided into train and test data in an 80:20 ratio. Then, the Random Forest classifier is used to generate a prediction for each sample. In our classifier, the number of estimators or trees in the forest was 100. 'Gini Impurity' is used as the function to measure the quality of the split. The probability of incorrect classification of a new instance of a random variable if that new instance were randomly categorized according to the distribution of class labels from the data set is known as 'Gini Impurity' [12]. A minimum of two samples are required to split an internal node, while one sample is required to be a leaf node. Nodes are expanded until there are no more than two samples in each leaf. It considers 10 features while looking for the best split and draws all the samples from training data to train each base estimator. Finally, the prediction result was stored in a separated column, and later, a feed-forward neural network was trained using it as additional input data.

*2) Neural Network:* A neural network is a set of algorithms that seek to link basic connections in a set of data in a cycle that mimics how the cerebrum portion of the human brain works. In other words, it's a computerized network of neurons that analyses data and generates output based on inputs.

The MLPClassifier of SKlearn [2] is implemented to develop our multilayer perceptron neural network model. The deep neural network model for this work has an input layer having 121 neurons (120 columns from the dataset and 1 additional column generated using RF classifier) and 4 hidden layers each having 256, 128, 64, and 32 neurons consecutively with a rectified linear unit(relu). Nonlinear activation functions cannot be utilized in networks with numerous layers due to the vanishing gradient problem. As a result, rectified linear activation is utilized as the default activation function for multilayer perceptron and convolutional neural networks. The Relu, therefore, fixes the Vanishing Gradient Problem. The output layer consists of 5 neurons having a softmax activation function for classifying the samples in 5 different classes. The probability of each class is given in the output layer and considered the one with the highest probability to be in our prediction class. 'Adam' solver, which is a stochastic gradient-based optimizer since it is the best fit for our large dataset in terms of training time and validation score. A batch size of 200 was considered for the stochastic optimizer. The starting learning rate was 0.001 and was maintained throughout the process. This learning rate controls the step size and updates the weights. 200 epochs are also used while training this model. To determine the appropriate number of epochs, accuracy vs epochs are observed. According to our observation, accuracy begins to saturate after running almost 200 epochs. 200 epochs are found to be optimal in this case as the error rate starts to increase after this count.

## IV. Dataset Description

The dataset that is used for this work is the CICMalDroid 2020 [13]. The target of this research is to develop a malware detection system for Android applications which will be efficient enough to detect malware in real-time. These were gathered from a variety of sources, including the VirusTotal service, the Contagio security blog, AMD, and other datasets from recent studies. The dataset comprises samples that were collected from 2017 to 2018 and are recent, sophisticated, and dynamic compared to other publicly available datasets. Therefore the dataset is divided into 5 distinct categories, which are Adware, Banking malware, SMS malware, Riskware, and Benign.

CopperDroid [13], a VMI-based dynamic analysis system, was used to dynamically analyze the data in order to automatically recreate low-level OS-specific and high-level Android-specific behaviors of Android samples. 13,077 out of 17,341 samples were successful, while the rest failed because of time-outs, invalid APK files, and memory allocation failures. While examining the samples of different groups we found 3,904 SMS malware which is almost 34% of the total samples. In addition, 2,546 Riskware were found which is nearly 22% of the total samples. Adware comprises 1253 which is almost 11% of the total samples. Banking has got 2,100 which is around 18% and finally Benign is 1,795 which is 15.5% approximately.

In CopperDroid, the APK files were inspected, and the run-time nature was recorded in log files. CopperDroid's output analysis findings are in JSON format for easy parsing and other auxiliary data. Statically extracted data such as intents, permissions, and services, frequency counts for various file types, obfuscation instances, and sensitive API invocations are among the study's results. The study comprised dynamically observed behaviors, which were split into three categories: system calls, binder calls, and composite behaviors.

All the 470 features in the dataset have been divided into three sections. Among these 7.4% of the data belong to the Composite Behaviors sector, 29.6% represents the API and System calls group and 63.0% of the data falls under the Binder Calls section.

## V. Experimentation

The CICMalDroid 2020 [13] dataset is used for this work . It consists of 11,598 rows and 470 columns in total. To improve the generalizability of our model data preprocessing was done. In order to get an idea of the data distribution, several tables and charts have been generated. Besides, standardization was performed to scale all values within a certain range for better prediction. The top 120 features were extracted out of the 470 features from the dataset during feature selection using weka and sklearn. The dataset was then divided into train and test data in an 80:20 ratio. A random forest classifier was then used to generate a prediction for each sample. The overall accuracy was 93.6%. The total training time for performing random forest was 2.1 seconds. To enhance overall accuracy, the predicted column is used as an additional input to train a feed-forward neural network. 200 epochs were used while training this model. The overall improved accuracy was 97.5%. The total training time for performing random forest and MLP was 22.945 seconds which is suitable to implement this model in real-time. The proposed model (Random Forest + MLP ) gave the highest accuracy of 0.98 and took 22.945 seconds to get trained. Random Forest and MLP alone have an accuracy of 0.94 and 0.93 respectively. SVM gave the least accuracy of 0.45.

The bar chart for representing the accuracy for all the models that have been used in this work is given below:

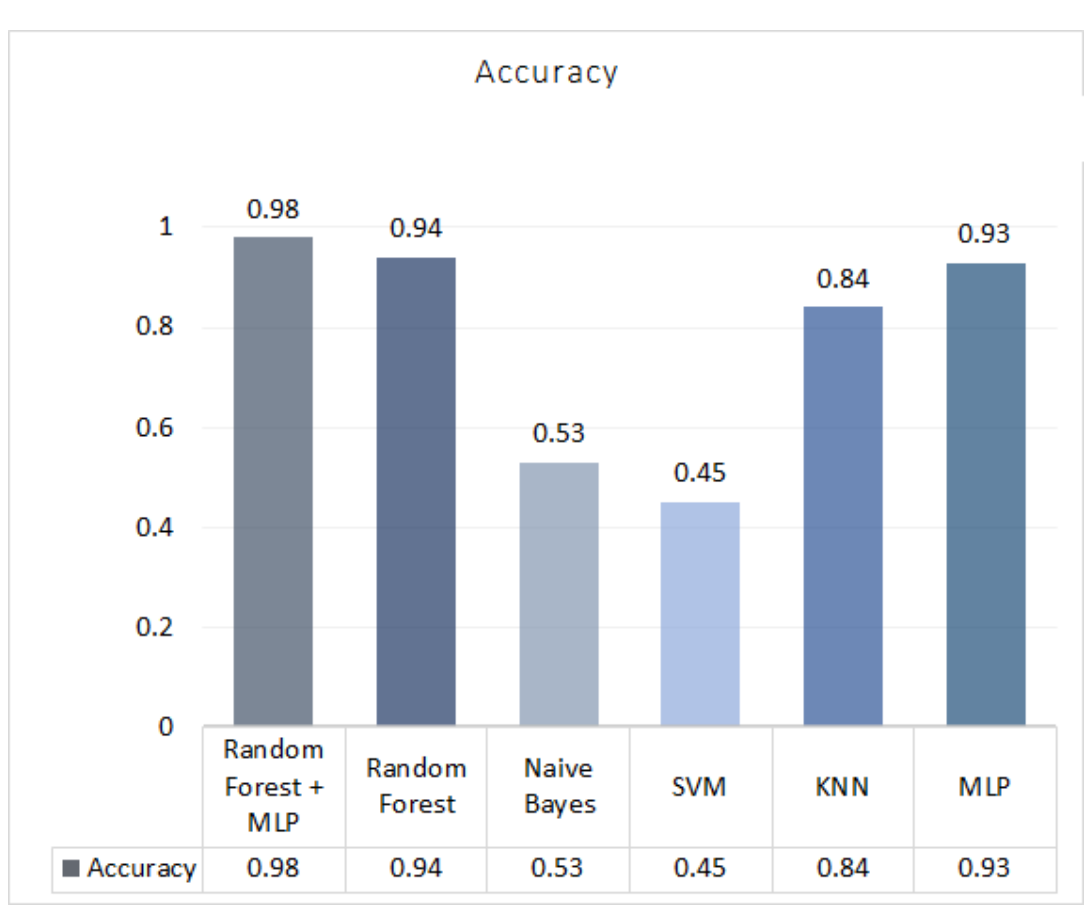

Fig. 2. Accuracy comparison

## VI. Result Analysis

An ensemble model of Random Forest and Multilayer perceptron network was proposed in this paper. However, various machine learning and neural network models such as Random Forest Classifier, Multilayer Perceptron Network, K Neighbour Classifier, Support Vector Machine, and Naive Bayes Classifier were used for comparative analysis. A confusion matrix was generated for each algorithm to represent the accuracy of each class. Furthermore, Precision, Recall, and F1-score were considered as evaluation metrics in this analysis. The accuracy, precision, recall, and f1-score of the test data are presented in the classification Report, which is a performance evaluation metric.

In our proposed method, the accuracy for Adware class was 0.97 or 97%, Banking class was 0.96 or 96%, Actual SMS was 0.99 or 99% (which was the highest), Riskware class was 0.96 or 96% and the Benign class was 0.97 or 97%. The confusion matrix of the proposed model is given below :

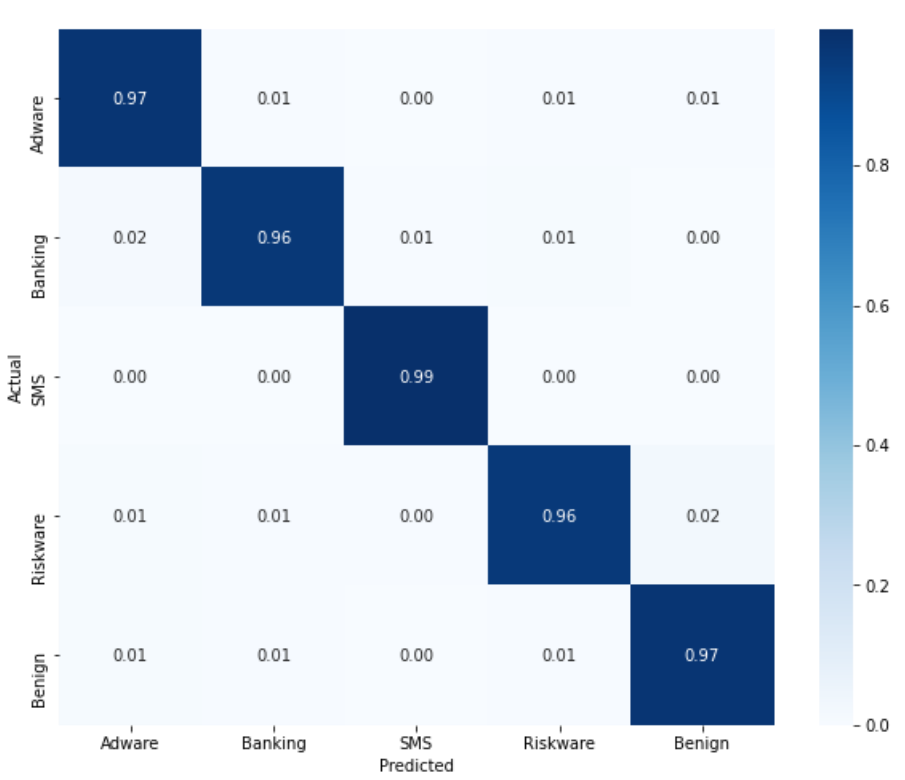

Fig. 3. Confusion matrix of the proposed model.

The precision value is highest for the SMS Malware and Mobile Riskware classes. The precision, recall, and f1-score values for all the classes are relatively very high and are above 90% compared to classification reports of other algorithms. The recall value and f1-score are highest for the SMS Malware class. The overall macro average of precision is 0.98, recall is 0.97 and f1-score is 0.98.

The precision value, recall value, and f1-score are highest for the SMS Malware class. However, precision and f1-score for the Adware class are comparatively lower than the rest of the class while using the Random Forest classifier. By implementing MLP, the precision value, recall value, and f1-score are highest for the SMS Malware class. However, precision, recall, and f1-score values for the Adware class are comparatively lower than the rest of the class while using Multilayer Perceptrons. When SVM was applied to the dataset, the precision value was highest for the Banking Malware class, recall value and f1-score were highest for the SMS Malware class. However, precision, recall, and f1-score for all the classes are comparatively lower while using SVM. The implementation of KNN on the dataset led to the precision value, recall value, and f1-score being highest for the SMS Malware class. However, precision, recall, and f1-score for all the classes are comparatively lower while using KNN. After the application of Naive Bayes on the dataset, the precision values are highest for the Mobile Riskware and Benign classes, recall value, and f1-score are highest for the SMS Malware class. However, precision, recall, and f1-score for all the classes are comparatively lower while using Naive Bayes. The table below shows the combined classification report for all models implemented. It consists of precision, recall, f1-score, support values, accuracy, macro average, and the weighted average for all the classes. From this table, it can be concluded that the proposed model i.e. the combined Random Forest and Multilayer Perceptrons (RF+MLP) model generates the highest scores compared to the rest.

TABLE I
Classification report of the proposed model (RF+MLP), Random Forest, MLP, SVM, KNN, and Naive Bayes

| Algorithm | Accuracy | F1-score | Precision | Recall |
|---|---|---|---|---|
| **Random Forest + MLP** | .98 | .97 | .98 | .97 |
| Random Forest | .94 | .93 | .93 | .93 |
| Naive Bayes | .53 | .45 | .44 | .57 |
| SVM | .45 | .31 | .62 | .34 |
| KNN | .84 | .80 | .81 | .80 |
| MLP | .93 | .91 | .91 | .91 |

Random Forest is an effective model for this work as it is capable of handling binary, category, and numerical features. Outliers were handled by Random Forest by effectively binning them. Random forest helped to reduce the overall error rate; allowing the larger classes to have a low error rate. The Random Forest technique was used to help this project handle large amounts of data with hundreds of variables. When a class is more infrequent than other classes in the data, it can automatically balance data sets. The approach also works quickly with variables, making it suited for more complex tasks like this one. On the other hand, MPL is ideal for similar classification tasks. The data in the datasets was numerical, and MLP works better with it. MLP helped with non-linear and complex situations. Finally, the proposed hybrid learning combining Random Forest and MLP helped in obtaining all these advantages and allowed to achieve higher performance, more flexibility, robustness, and accuracy with less computational complexity. Hybridizing the algorithm (Random Forest with MLP), which is our proposed method, the improved accuracy for Adware class was increased to 97%, Banking class was 96%, Actual SMS was 99% (which was the highest), Riskware class was 96%, and the Benign class was 97%. The higher correlation from the additional input helped to achieve better results, which was the goal of using this proposed model. However, SVM failed due to a heterogeneous dataset and numerous independent features. Similarly, Naive-Bayes' classifier underperformed in this work.

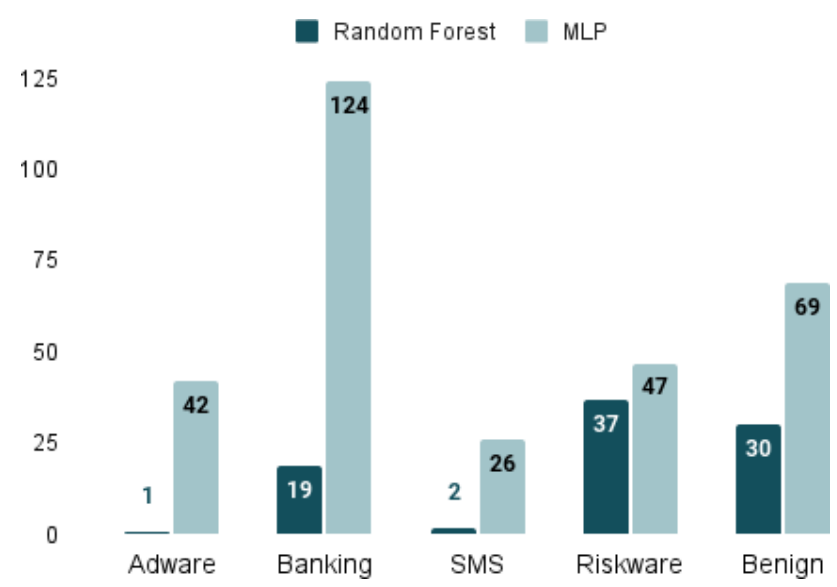

Fig. 4. incorrect predictions by RF and MLP alone that are successfully identified by the hybrid model

The above figure demonstrates the number of samples that have been detected correctly by our hybrid model but incorrectly identified by Random Forest and MLP alone. A total of 89 samples have been found to be successfully identified by the proposed hybrid model where Random Forest alone failed to predict correctly. Similarly, MLP alone misidentifies a total of 308 samples, where the majority were Banking malware (124 samples), which have been detected correctly by the hybrid model. Moreover, the predictions of several Riskware samples have been found to be inaccurate by both Random Forest and MLP alone, but the proposed hybrid model is able to identify those effectively. The issue with using an unbalanced dataset to train a model is that it will only be biased towards the majority class. This creates a problem when we are interested in the prediction of the minority class. Thus, to mitigate these possible problems, we have used a stratified train-test split that divides the dataset into train and test sets while maintaining the same proportions of samples in each class as the original dataset.

## VII. Conclusion and Future work

Malware detection has become an indispensable part of mobile devices for network security and privacy. This is because malware is advancing and becoming stronger day by day alongside the continuous upsurge of mobile users, especially Android mobile phone users. Although intrusion detection is evolving rapidly to protect privacy and electronic devices, the vandals are also progressing at a fast rate which is causing jeopardy in network security. Therefore, more efficient security measures are becoming a dire need. Moreover, advanced malware is finding its way into personal devices faster than before and affecting them coherently for which existing intrusion detection systems are not being able to pace up due to being slow and inefficient. Real-time malware detection can be beneficial in this sector as it is dynamic and fast but not much research was conducted in this field. This is where our research comes in as an extra coat of security by adding more promising features to the existing layers of security and filling the void in the research field of real-time malware detection. This paper can be improved for future endeavors by making the application built dynamic. Furthermore, this work only detects malware in Android applications which can also be extended for detecting malware in applications supported by the iOS platform using proper dataset in the future.

## References

[1] R. Feng, J. Q. Lim, S. Chen, S.-W. Lin, and Y. Liu, "Seqmobile: A sequence based efficient android malware detection system using rnn on mobile devices," *arXiv preprint arXiv:2011.05218*, 2020.

[2] B. Hendricks, *Study.com — Take Online Courses. Earn College Credit. Research Schools, Degrees Careers*. [Online]. Available: https://study.com/academy/lesson/Android-malware-infection-spread-impact.html.

[3] D. Sgandurra, L. Mun˜oz-Gonza´lez, R. Mohsen, and E. C. Lupu, "Automated dynamic analysis of ransomware: Benefits, limitations and use for detection," *arXiv preprint arXiv:1609.03020*, 2016.

[4] D. Palmer, *Mobile malware attacks are booming in 2019: These are the most common threats*, Jul. 2019. [Online]. Available: https : / / www. zdnet . com / article / mobile - malware - attacks - are - booming - in - 2019 - these - arehttps : / / www . overleaf . com / project / 612e131113dfbd3a4cab3f79 - the - most - common - threats/.

[5] Statista, *Android smartphone users in the United States 2014-2022*, Mar. 2021. [Online]. Available: https : / / www . statista . com / statistics / 232786 / forecast - of - andrioid-users-in-the-us/.

[6] J. Callaham, *The history of Android: The evolution of the biggest mobile OS in the world*, May 2021. [Online]. Available: https://www.Androidauthority.com/history-Android-os-name-789433/.

[7] Y. Xin, L. Kong, Z. Liu, Y. Chen, Y. Li, H. Zhu, M. Gao, H. Hou, and C. Wang, "Machine learning and deep learning methods for cybersecurity," *Ieee access*, vol. 6, pp. 35 365–35 381, 2018.

[8] N. Varol, A. F. Aydogan, and A. Varol, "Cyber attacks targeting android cellphones," in *2017 5th International Symposium on Digital Forensic and Security (ISDFS)*, IEEE, 2017, pp. 1–5.

[9] M. Alazab, M. Alazab, A. Shalaginov, A. Mesleh, and A. Awajan, “Intelligent mobile malware detection using permission requests and api calls,” *Future Generation Computer Systems*, vol. 107, pp. 509–521, 2020.

[10] W. Wang, M. Zhao, and J. Wang, “Effective android malware detection with a hybrid model based on deep autoencoder and convolutional neural network,” *Journal of Ambient Intelligence and Humanized Computing*, vol. 10, no. 8, pp. 3035–3043, 2019.

[11] L. Sthle and S. Wold, “Analysis of variance (anova),” *Chemometrics and Intelligent Laboratory Systems*, vol. 6, no. 4, pp. 259–272, 1989, ISSN: 0169-7439. DOI: https://doi.org/10.1016/0169-7439(89)80095-4. [Online]. Available: https://www.sciencedirect.com/science/article/pii/0169743989800954.

[12] B. Ambielli, *Gini Impurity (With Examples)*, Oct. 2017. [Online]. Available: https://bambielli.com/til/2017-10-29-gini-impurity/.

[13] S. Mahdavifar, A. F. A. Kadir, R. Fatemi, D. Alhadidi, and A. A. Ghorbani, “Dynamic android malware category classification using semi-supervised deep learning,” in *2020 IEEE Intl Conf on Dependable, Autonomic and Secure Computing, Intl Conf on Pervasive Intelligence and Computing, Intl Conf on Cloud and Big Data Computing, Intl Conf on Cyber Science and Technology Congress (DASC/PiCom/CBDCom/CyberSciTech)*, IEEE, 2020, pp. 515–522.